\documentclass[twocolumn,showpacs,preprintnumbers,amsmath,amssymb]{revtex4}

\usepackage{graphicx}
\usepackage{dcolumn}
\usepackage{bm}

\def\Journal#1#2#3#4{{#1} {\bf #2}, #3 (#4)}
\def\JPG{Journal of Phys. G} 
\def\ARNPS{Annu. Rev. Nucl. Part. Sci.} 
\def\ANP{Ann. Phys.}

\def\APJS{Astrophys. J. Suppl}

\def\IJMPE{Int. J. Mod. Phys. E}
\def\JHEP{JHEP}

\def\MPLA{Mod. Phys. Lett. A}

\def\NPB{Nucl. Phys. B}

\def\PLB{{Phys. Lett.} B}

\def\PRL{Phys. Rev. Lett.}
\def\PRD{Phys. Rev. D}

\def\SJNP{Sov. J. Nucl. Phys.}

\def\PAN{Phys. Atom. Nucl.}

\begin{document}


\title{Remark on the minimal seesaw model and leptogenesis \\ with tri/bi-maximal mixing}\author{Teruyuki Kitabayashi}
\email{teruyuki@keyaki.cc.u-tokai.ac.jp}
\affiliation{
Department of Physics, Tokai University,
1117 Kitakaname, Hiratsuka, Kanagawa, 259-1292, Japan
}
\date{\today}

\begin{abstract}
We have studied a leptogenesis scenario in the framework of the minimal seesaw model with tri/bi-maximal mixing. Usually, at least one of the elements in the Dirac mass matrix is fixed to be zero, for example, we denote it by $b_2=0$. We have pointed out that the absolute value of the CP asymmetry has several minimums and maximums with non-zero $b_2$. Thus one can expect that more rich phenomena, such as an enhanced leptogenesis, are hidden in the $b_2 \neq 0$ space. 
\end{abstract}


\maketitle

\section{\label{sec:Introduction}Introduction}
The recent neutrino oscillation experiments have provided us with robust evidence that neutrinos have tiny masses and their flavor mixing involves two large angles and one small angle \cite{experiments}. A global analysis of current neutrino oscillation data  yields $7.2\times 10^{-5}$eV$^2$ $\le \Delta m_s^2 \le 8.9\times 10^{-5}$eV$^2$ and $1.7\times 10^{-3}$eV$^2$ $\le \Delta m_a^2 \le 3.3\times 10^{-3}$eV$^2$ for the squared mass differences of solar and atmospheric neutrinos and $30^\circ \le \theta_{12} \le 38^\circ$, $36^\circ \le \theta_{23}\le 54^\circ$ and $0^\circ\le\theta_{13} < 10^\circ$ for the flavor mixing angles at the $99\%$ confidence level (the best-fit values are $\Delta m_s^2 = 8.0\times 10^{-5}$eV$^2$, $\Delta m_a^2 = 2.5\times 10^{-3}$eV$^2$, $\theta_{12} = 34^\circ$, $\theta_{23}=45^\circ$ and $\theta_{13}=0^\circ$)\cite{Strumia2005}. Where we define $\Delta m_s^2 = m_2^2 - m_1^2$ and $\Delta m_a^2 = \vert m_3^2 - m_2^2 \vert$ with the neutrino mass eigenvalues $m_1,m_2$ and $m_3$. 

There are two important questions: (1) why are the two neutrino mixing angles large, almost maximal, and is the one angle small? and (2) why are the neutrino masses so tiny? Tri/bi-maximal mixing is the one of the lepton mixing ansatz to answer the first question, which is characterized by the following mixing matrix \cite{Tribimaximal}
\begin{eqnarray}
U=\left(
  \begin{array}{ccc}
    \sqrt{\frac{2}{3}}  & \frac{1}{\sqrt{3}} & 0 \\
    -\frac{1}{\sqrt{6}} & \frac{1}{\sqrt{3}} & -\frac{1}{\sqrt{2}}\\
    -\frac{1}{\sqrt{6}} & \frac{1}{\sqrt{3}} & \frac{1}{\sqrt{2}}\\
  \end{array}
\right).
\end{eqnarray}
The texture of this matrix consists of $\sin\theta_{23} = 1/\sqrt{2}$, $\sin\theta_{12}=1/\sqrt{3}$ and $\sin\theta_{13}=0$. Thus, the tri/bi-maximal mixing matrix could approximately describe the best fit values of the current neutrino oscillation data: $\theta_{23}=45^\circ$, $\theta_{12} = 34^\circ$ and $\theta_{13}=0^\circ$. Recently, many theoretical efforts have been made to produce the tri/bi-maximal mixing pattern arising from some relevant symmetries\cite{Tribimaximal_symmetry}.

On the other hand, an attractive idea to answer the second question is given by the seesaw mechanism \cite{seesaw} which explains the smallness of the neutrino masses through the existence of right-handed Majorana neutrinos. Among many realistic seesaw models existing in the literature, the minimal seesaw model contains only two right-handed Majorana neutrinos, thus it is the most economical one \cite{minimalSeesaw}. 
Also, the seesaw mechanism provides a very natural explanation of the baryon asymmetry in the Universe through the baryogenesis via leptogenesis scenarios \cite{leptogenesis}. The baryon asymmetry has been measured by the Wilkinson Microwave Anisotropy Probe (WMAP) as a baryon-photon ratio $\eta_B = (6.1_{-0.2}^{+0.3})\times 10^{-10}$ \cite{WMAP} . 

The leptogenesis scenarios in the minimal seesaw model have been studied by many groups \cite{minimalSeesawAndLeptogenesis}. However, as mentioned below, if we try to analytically solve the Dirac neutrino mass matrix $m_D$ in terms of the observed element of the light neutrino mass matrix, it is sometime assumed that at least one of the elements in  $m_D$ should be fixed to a constant or we should assume an equality between two or more entries in the matrix $m_D$. For example, some groups have estimated the baryon asymmetry with the following Dirac mass matrix
\begin{eqnarray}
m_D
=\left(
  \begin{array}{cc}
    A_1 & B_1  \\
    A_2 & B_2  \\
    A_3 & B_3  \\
  \end{array}
\right)
=
\left(
  \begin{array}{cc}
    \ast & \ast  \\
    \ast & 0     \\
    \ast & \ast  \\
  \end{array}
\right),
\end{eqnarray}
which is so-called texture one-zero mass matrix. However, it is a natural question, what happen if the $B_2$ element deviates from zero? This is our question in this paper. 

\section{\label{sec:MinimalSeesawModel}Minimal seesaw model}
We show the brief review of the minimal seesaw model. The minimal seesaw Lagrangian can be written as 
\begin{eqnarray}
L=-\overline{\ell}_LM_L\ell_R - \overline{\nu}_Lm_DN_R + \frac{1}{2}\overline{N}_R^c M_RN_R + h.c.,
\end{eqnarray}
where $\nu_L = (\nu_e,\nu_\mu,\nu_\tau)^T$, $\ell_L =(e, \mu, \tau)^T$ and $N_R=(N_1,N_2)^T$ denote the left-handed (light) neutrinos, the left-handed charged leptons and the right-handed (heavy) neutrinos, respectively. We have assumed that the mass matrices of both the heavy neutrinos $M_R=\textrm{diag}.(M_1,M_2)$ and the charged lepton $M_\ell=\textrm{diag}.(m_e,m_\mu,m_\tau)$ are diagonal and real. The Dirac neutrino mass matrix $m_D$ is a $3 \times 2$ rectangular matrix, which can be written by 6 parameters ($a_1, a_2, a_3, b_1, b_2, b_3$) as \cite{tribi_leptogenesis}
\begin{eqnarray}
m_D = 
\left(
  \begin{array}{cc}
    \sqrt{M_1}a_1  & \sqrt{M_2}b_1   \\
    \sqrt{M_1}a_2  & \sqrt{M_2}b_2   \\
    \sqrt{M_1}a_3  & \sqrt{M_2}b_3   \\
  \end{array}
\right).
\end{eqnarray}

Via the well-known seesaw mechanism, we can obtain the symmetric $3 \times 3$ Majorana mass matrix for light neutrinos as $M_\nu = -m_D M_R^{-1} m_D^T$ and which yields the following equation
\begin{eqnarray}
M_\nu
&\equiv&
\left(
  \begin{array}{ccc}
    M_{11} & M_{12} & M_{13} \\
           & M_{22} & M_{23} \\
           &        & M_{33} \\
  \end{array}
\right)
\nonumber\\
&=&
\left(
  \begin{array}{ccc}
    a_1^2 + b_1^2   & a_1a_2 + b_1b_2   &  a_1a_3 + b_1b_3  \\
                    & a_2^2 + b_2^2     &  a_2a_3 + b_2b_3  \\
                    &                   &  a_3^2 + b_3^2   \\
  \end{array}
\right).
\nonumber\\
\label{Eq:Mnu}
\end{eqnarray}
This equation contains 6 parameters, however, we have only 5 independent condition, the sixth condition is redundant because of $\textrm{det}(M_\nu)=0$ \cite{minimalSeesaw}. Therefore, if we choose one of the elements in the Dirac mass matrix $(a_1, a_2, a_3, b_1, b_2, b_3)$ as a constant term, we can analytically solve 5 parameters out of 6 of $m_D$ in terms of the element of the light neutrino mass matrix. For $M_{11} \neq 0$, the solution in terms of $a_1$ or $b_1$ is 
\begin{eqnarray}
a_1 = \sqrt{M_{11}-b_1^2}, \quad or \quad b_1=\sqrt{M_{11}-a_1^2},
\nonumber
\end{eqnarray}
\begin{eqnarray}
a_2 = \frac{1}{M_{11}}\left[ a_1M_{12} - \sigma_2b_1\sqrt{M_{11}M_{22}-M_{12}^2}  \right],
\nonumber
\end{eqnarray}
\begin{eqnarray}
a_3 = \frac{1}{M_{11}}\left[ a_1M_{13} - \sigma_3b_1\sqrt{M_{11}M_{33}-M_{13}^2}\right],
\nonumber
\end{eqnarray}
\begin{eqnarray}
b_2 = \frac{1}{M_{11}}\left[ b_1M_{12} + \sigma_2a_1\sqrt{M_{11}M_{22}-M_{12}^2}
\right],
\nonumber
\end{eqnarray}
\begin{eqnarray}
b_3 = \frac{1}{M_{11}}\left[ b_1M_{13} + \sigma_3a_1\sqrt{M_{11}M_{33}-M_{13}^2}\right],
\nonumber
\end{eqnarray}
\begin{eqnarray}
M_{23} &=& \frac{1}{M_{11}} \left[ M_{12}M_{13} \right.
\nonumber\\ 
 &+& \left.\sigma_2\sigma_3\sqrt{(M_{11}M_{22}-M_{12}^2)(M_{11}M_{33}-M_{13}^2)} \right],
\nonumber\\
\label{Eq:reconstruction}
\end{eqnarray}
where $\sigma_{2,3}=\pm 1$ and $\textrm{sgn}(a_1)=+1$ \cite{tribi_leptogenesis, reconstruction}. The last equation is a consistency condition due to $\textrm{det}(M_\nu)=0$ and we can use it to fix the sign of $\sigma_2\sigma_3$. The solution in terms of $a_2$ or $b_2$ is obtained under the exchanges $a_1 \leftrightarrow a_2$, $b_1 \leftrightarrow b_2$, $M_{11} \leftrightarrow M_{22}$, $M_{13} \leftrightarrow M_{23}$, $\sigma_1 \leftrightarrow \sigma_2$, for $M_{22}\neq 0$ and $\textrm{sgn}(a_2)=+1$ with $\sigma_{1,3}=\pm 1$. Similarly, under the transformations $a_1 \leftrightarrow a_3$, $b_1 \leftrightarrow b_3$, $M_{11} \leftrightarrow M_{33}$, $M_{12} \leftrightarrow M_{23}$, $\sigma_1 \leftrightarrow \sigma_3$ with $\sigma_{2,1}=\pm 1$, we obtain the solutions in terms of $a_3$ or $b_3$ for $M_{33}\neq 0$ and $\textrm{sgn}(a_3)=+1$. 

The light neutrino mass matrix $M_\nu$ is also calculated to be $M_\nu = U^\ast M^{\textrm{diag}} U^\dagger$ where $M^{\textrm{diag}}=\textrm{diag}.(m_1,m_2,m_3)$ is the diagonal neutrino mass matrix and $U$ is the Maki-Nakagawa-Sakata mixing matrix which can be parameterized by
\begin{widetext}
\begin{equation}
U=\left(
  \begin{array}{ccc}
    c_{12}c_{13}    & s_{12}c_{13} & s_{13}e^{-i\delta} \\
    -s_{12}c_{23} - c_{12}s_{23}s_{13}e^{i\delta} 
        & c_{12}c_{23} - s_{12}s_{23}s_{13}e^{i\delta}  
        & s_{23}c_{13}\\
    s_{12}s_{23} - c_{12}c_{23}s_{13}e^{i\delta} 
        & -c_{12}s_{23} - s_{12}c_{23}s_{13}e^{i\delta}  
        & c_{23}c_{13}\\
  \end{array}
\right)
\left(
  \begin{array}{ccc}
    e^{i\alpha_1/2} & 0               & 0 \\
    0               & e^{i\alpha_2/2} & 0 \\
    0               & 0               & 1 \\
  \end{array}
\right),
\label{Eq:UMNS}
\end{equation}
\end{widetext}
where $c_{ij}=\cos\theta_{ij}$ and $s_{ij}=\sin\theta_{ij}$ $(i,j = 1,2,3)$. There are two types of the leptonic CP violation phases: one is the Dirac phase $\delta$ and the other are the two Majorana phases $\alpha_1, \alpha_2$ if neutrinos are Majorana particle \cite{PDG}. Now, we can reconstruct the Dirac mass matrix elements $(a_1, a_2, a_3, b_1, b_2, b_3)$ in term of the low energy neutrino observables $m_1, m_2, m_3, \theta_{12}, \theta_{23}, \theta_{13}$ and CP violating phases $\delta, \alpha_1, \alpha_2$ with heavy neutrino masses $M_1, M_2$. 

Moreover, in the framework of the minimal seesaw model, $\textrm{det}(M_\nu)=0$ condition holds if one of the neutrino mass eigenvalue $(m_1,m_2,m_3)$ is exactly zero \cite{minimalSeesaw}. As a result, the experimental results of the square mass differences can be explained by the two types of mass hierarchies. One is called normal mass hierarchy: $(m_1,m_2,m_3) = \left(0, \sqrt{\Delta m_s^2}, \sqrt{\Delta m_s^2 + \Delta m_a^2}\right)$, and the other is called the inverted mass hierarchy: $(m_1,m_2,m_3) = \left(\sqrt{\Delta m_a^2 - \Delta m_s^2}, \sqrt{\Delta m_a^2}, 0\right)$.

\section{\label{sec:Leptogenesis}Leptogenesis}
We give an outline of the leptogenesis. The baryon-photon ratio $\eta_B$ can be translated into the baryon asymmetry $Y_B = (n_B-n_{\overline{B}})/s$ as $\eta_B = 7.04Y_B$, and in leptogenesis scenario, this baryon asymmetry produced via the sphaleron process is related to the lepton asymmetry $Y_L$ \cite{lepto_reviews}:
\begin{eqnarray}
Y_B= -\frac{8N+4m}{14N+9m} Y_L,
\end{eqnarray}
where $N$ is the number of generation of fermions and $m$ is the number of Higgs doublets. In the particle contents of the standard model, $N=3$ and $m=1$ thus $Y_B \simeq -0.549Y_L$. 

The lepton asymmetry $Y_L$ is parameterized by three term, the dilution factor $d$, the CP-asymmetry parameter $\epsilon$ and the effective number of relativistic degree of freedom $g_\ast$
\begin{eqnarray}
Y_L = d\frac{\epsilon}{g_\ast}.
\end{eqnarray}
where, for instance, we have $g_\ast \simeq 106.75$ in the standard model.
The dilution factor $d$ describes the wash-out of the lepton asymmetry due to the various lepton number violating processes. We should solve the full Boltzmann equations to determine the magnitude of the dilution factor exactly; however, we can use good analytical approximation by Nielsen and Takanishi in Ref.~\cite{Nielsen01}
\begin{eqnarray}
d \sim \frac{1}{2.0 \sqrt{K^2+9}},
\end{eqnarray}
for $0 \le K \le 10$ and
\begin{eqnarray}
d \sim \frac{0.3}{K(\ln K)^{0.6}},
\label{Eq:dilution}
\end{eqnarray}
for $10 \le K \le 10^6$, where
\begin{eqnarray}
K &=& \frac{M_{pl}}{1.66\sqrt{g_\ast}(8\pi v^2)}\frac{(m_D^\dagger m_D)_{11}}{M_1}
\nonumber \\
&\simeq& \frac{1}{10^{-3}eV}\left(\vert a_1 \vert^2 + \vert a_2 \vert^2 + \vert a_3 \vert^2\right).
\label{Eq:K}
\end{eqnarray}
with the Planck mass $M_{pl} \simeq 1.22\times 10^{19}$ GeV and the vacuum expectation value of the Higgs particle $v \simeq 174$ GeV.

The CP asymmetry $\epsilon$ have generated by the decay processes of the heavy Majorana neutrino:
\begin{eqnarray}
\epsilon=\sum_{i=1,2}\epsilon_i=\sum_{i=1,2}
\frac{\Gamma(N_i\rightarrow \ell H) - \Gamma(N_i\rightarrow \bar{\ell} H^\ast)}
{\Gamma(N_i\rightarrow \ell H) + \Gamma(N_i\rightarrow \bar{\ell} H^\ast)},
\end{eqnarray}
where $H$ denotes the Higgs doublet in the standard model. 

Now, we assume a hierarchical mass pattern of the heavy neutrinos $M_1 \ll M_2$. In this case, the interactions of $N_1$ can be in thermal equilibrium when $N_2$ decays and the asymmetry caused by the $N_2$ decay is washed out by the lepton number violating processes with $N_1$. Thus, only the decays of $N_1$ are relevant for generation of the final lepton asymmetry $\epsilon \simeq \epsilon_1$. In this case, the CP asymmetry parameter in the minimal seesaw model calculated to be \cite{minimalSeesawAndLeptogenesis}:
\begin{eqnarray}
\epsilon &=& \frac{1}{8\pi v^2} \frac{\textrm{Im}\left[(m_D^\dagger m_D)_{12}^2 \right]}{(m_D^\dagger m_D)_{11}} f\left(\frac{M_2}{M_1} \right)
\nonumber\\
&=&\frac{M_2}{8\pi v^2}
\frac{\textrm{Im}\left[(a_1^\ast b_1 + a_2^\ast b_2 + a_3^\ast b_3)^2 \right]}{\vert a_1 \vert^2 + \vert a_2 \vert^2 + \vert a_3 \vert^2}f\left(\frac{M_2}{M_1} \right),
\nonumber\\
\end{eqnarray}
where the function $f(x)$ is given by
\begin{eqnarray}
f(x) = x\left[1 - (1+x^2)\ln\left(\frac{1+x^2}{x^2}\right) + \frac{1}{1-x^2}\right].
\end{eqnarray}

\section{\label{sec:b2dependenceToEtaB}$b_2$ dependence to $\eta_B$}
As we have already addressed, the leptogenesis scenarios in the minimal seesaw model has been studied many groups in the case that at least one of the elements in the Dirac neutrino mass matrix are fixed to zero, such as the $b_2=0$ case:
\begin{eqnarray}
m_D = 
\left(
  \begin{array}{cc}
    \sqrt{M_1}a_1  & \sqrt{M_2}b_1   \\
    \sqrt{M_1}a_2  & 0   \\
    \sqrt{M_1}a_3  & \sqrt{M_2}b_3   \\
  \end{array}
\right).
\end{eqnarray}
Then, from Eq.~(\ref{Eq:Mnu}), the low energy neutrino mass matrix $M_\nu$ can be written by five parameters ($a_1, a_2, a_3, b_1, b_3$) as
\begin{eqnarray}
\left(
  \begin{array}{ccc}
    a_1^2 + b_1^2   & a_1a_2   &  a_1a_3 + b_1b_3  \\
                    & a_2^2    &  a_2a_3 \\
                    &          &  a_3^2 + b_3^2   \\
  \end{array}
\right)
\end{eqnarray}
and we can analytically explain these five parameters in terms of the five independent elements in the low energy neutrino mass matrix.  

However, to solve the Dirac mass matrix we can take not only $b_2=0$ but also $b_2 = X$ where $X$ is an any constant.  
\begin{eqnarray}
m_D = 
\left(
  \begin{array}{cc}
    \sqrt{M_1}a_1  & \sqrt{M_2}b_1   \\
    \sqrt{M_1}a_2  & \sqrt{M_2}X   \\
    \sqrt{M_1}a_3  & \sqrt{M_2}b_3   \\
  \end{array}
\right).
\end{eqnarray}
There are six non-vanishing entries but there are only five independent parameters. Then, the low energy neutrino mass matrix can be written by five parameters ($a_1, a_2, a_3, b_1, b_3$) as
\begin{eqnarray}
\left(
  \begin{array}{ccc}
    a_1^2 + b_1^2   & a_1a_2 + b_1X   &  a_1a_3 + b_1b_3  \\
                    & a_2^2 +X^2     &  a_2a_3 + Xb_3  \\
                    &              &  a_3^2 + b_3^2   \\
  \end{array}
\right).
\end{eqnarray}
This is same situation of $b_2=0$ case and we can analytically obtain these five parameters in terms of the five independent elements in $M_\nu$ again. Thus, $b_2 = 0$ is just special case of the $b_2=X$, e.g, both $b_2=0$ case and $b_2 \neq 0$ case such as $b_2 = 1, 2, 3,\cdots $ are in same ground for reconstruction of the Dirac mass matrix elements in term of the low energy observables. 

In this section, we ask the question, what happen if the $b_2$ element deviate from zero? To answer this question, we estimate the $b_2$ dependence to the baryon-photon ratio $\eta_B$ by numerical calculations. 

First, we choose the $b_2$ range. The solution of the elements in the Dirac mass matrix in the term of $b_2$ is obtained as $a_2 = \sqrt{M_{22}-b_2^2}$ etc. Thus, we take $b_2 = r \sqrt{M_{22}}$ with a parameter $r$ and use the following parameterization:
\begin{eqnarray}
a_2 &=& \sqrt{M_{22}-b_2^2}, \quad \quad b_2 = r \sqrt{M_{22}},
\nonumber\\
a_1 &=& \frac{1}{M_{22}}\left[ a_2M_{12} - \sigma_1b_2\sqrt{M_{22}M_{11}-M_{12}^2}  \right],
\nonumber\\
a_3 &=& \frac{1}{M_{22}}\left[ a_2M_{23} - \sigma_3b_2\sqrt{M_{22}M_{33}-M_{23}^2}\right],
\nonumber\\
b_1 &=& \frac{1}{M_{22}}\left[ b_2M_{12} + \sigma_1a_2\sqrt{M_{22}M_{11}-M_{12}^2}
\right],
\nonumber\\
b_3 &=& \frac{1}{M_{22}}\left[ b_2M_{23} + \sigma_3a_2\sqrt{M_{22}M_{33}-M_{23}^2}\right],
\nonumber\\
M_{13} &=& \frac{1}{M_{22}} \left[ M_{12}M_{23} \right.
\nonumber\\
&+& \left.\sigma_1\sigma_3\sqrt{(M_{22}M_{11}-M_{12}^2)(M_{22}M_{33}-M_{23}^2)} \right].
\nonumber\\
\end{eqnarray}
The parameter $r$ is to be varied in the range of $0 \le r \le 1$, where $r=0$ means the $b_2 = 0$ case while $r = 1$ means the $a_2 = 0$ case. 

For mixing angles, we assume the tri/bi-maximal mixing pattern, e.g., $\sin\theta_{23} = 1/\sqrt{2}$, $\sin\theta_{12}=1/\sqrt{3}$ and $\sin\theta_{13}=0$ to be consistent with the observed properties of neutrinos. Thus, in our analysis, the Dirac CP phase $\delta$ will not be active.  Also, we assume the light neutrinos have the normal hierarchical mass pattern with $\Delta m_s^2 = 7.0\times10^{-5}$eV$^2$ and $\Delta m_a^2 = 2.5\times 10^{-3}$ eV$^2$. In this case, only one Majorana phase $\alpha_2$ in Eq.~(\ref{Eq:UMNS}) is physically relevant. Finally, we assume the magnitude of the heaver neutrino mass $M_2 = 10M_1$ and use $g_\ast = 106.75$.

We show the numerical result in the figures. The main result in this study is shown in Fig.~\ref{fig:fig5} and Fig.~\ref{fig:fig7}. Other figures are shown to explain the behavior of the main result or to check the consistency of our analysis.

\begin{figure}
  \includegraphics{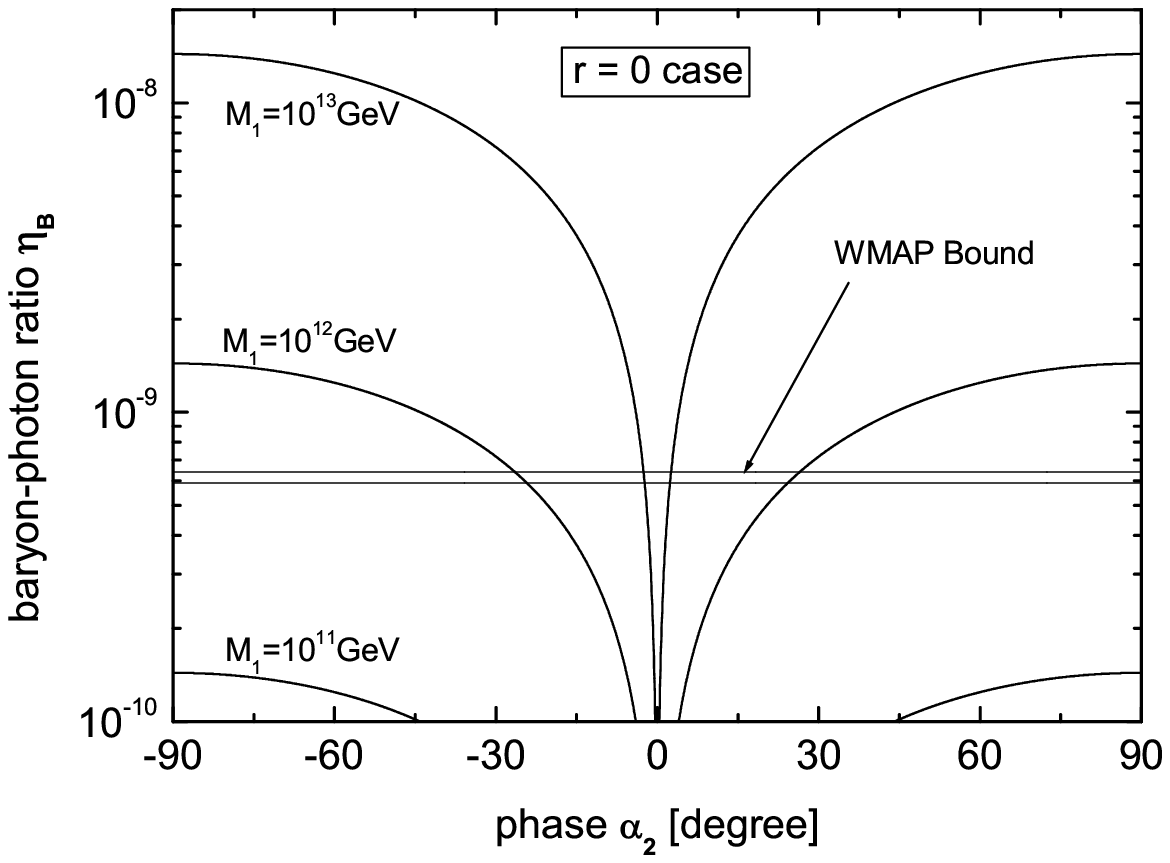}
  \caption{\label{fig:fig1}Baryon-photon ratio $\eta_B$ as a function of the Majorana phase $\alpha_2$ in the $r=0$ ($b_2 = 0$) case with various values of $M_1$. The imposed horizontal lines show the upper and lower bound of $\eta_B$ from the WMAP observation.}
\end{figure}

Fig.~\ref{fig:fig1} shows the $M_1$ dependence to the baryon-photon ratio $\eta_B$ in the case of $r = 0$, or equivalently $b_2=0$. This $r=0$ case was first estimated by Chang, Kang and Siyeon in Ref.~\cite{tribi_leptogenesis}. Our result is consistent with the Chang's solution and we have also confirmed that the baryon-photon ratio $\eta_B$ increases with the mass of heavy neutrino $M_1$. This $M_1$ dependence to the baryon-photon ratio $\eta_B$ can be understand by the following fact. The function $f(x) \sim -3/(2x)$ for large value of $x= M_2/M_1$, thus the CP asymmetry parameter $\epsilon$ becomes
\begin{eqnarray}
\epsilon &\sim& \frac{3}{16\pi v^2} M_1 \tilde{\epsilon},
\nonumber\\
\tilde{\epsilon} &\equiv& \frac{\textrm{Im}\left[(a_1^\ast b_1 + a_2^\ast b_2 + a_3^\ast b_3)^2 \right]}{\vert a_1 \vert^2 + \vert a_2 \vert^2 + \vert a_3 \vert^2}.
\label{Eq:epsilon_tild}
\end{eqnarray}
Since $\tilde{\epsilon}$ is independent of $M_1$, $\epsilon$ is nearly proportional to the mass of heavy neutrino $\epsilon \propto M_1$. 

\begin{figure}
  \includegraphics{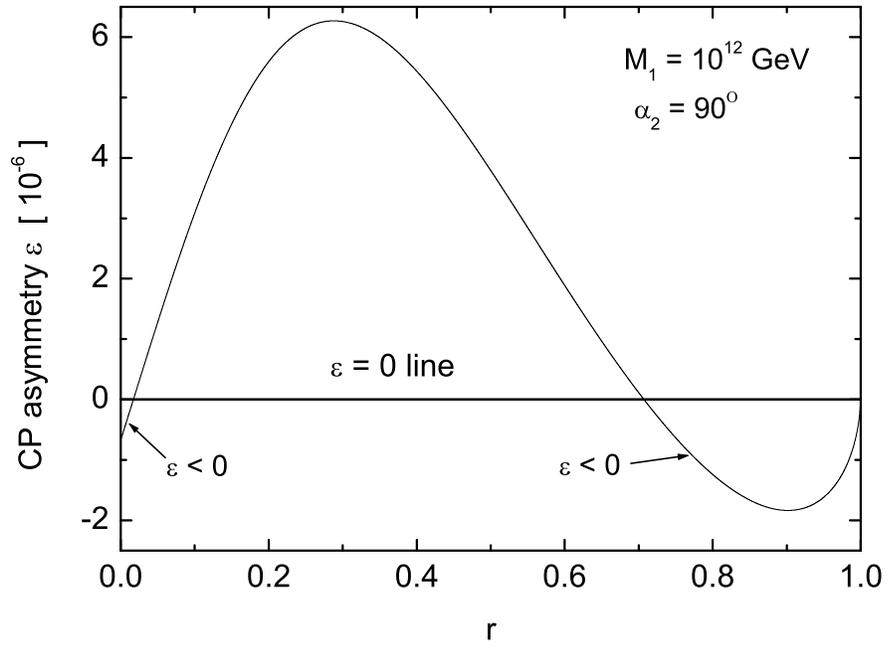}
  \caption{\label{fig:fig2}CP asymmetry parameter $\epsilon$ as a function of $r$.}
\end{figure}

\begin{figure}
  \includegraphics{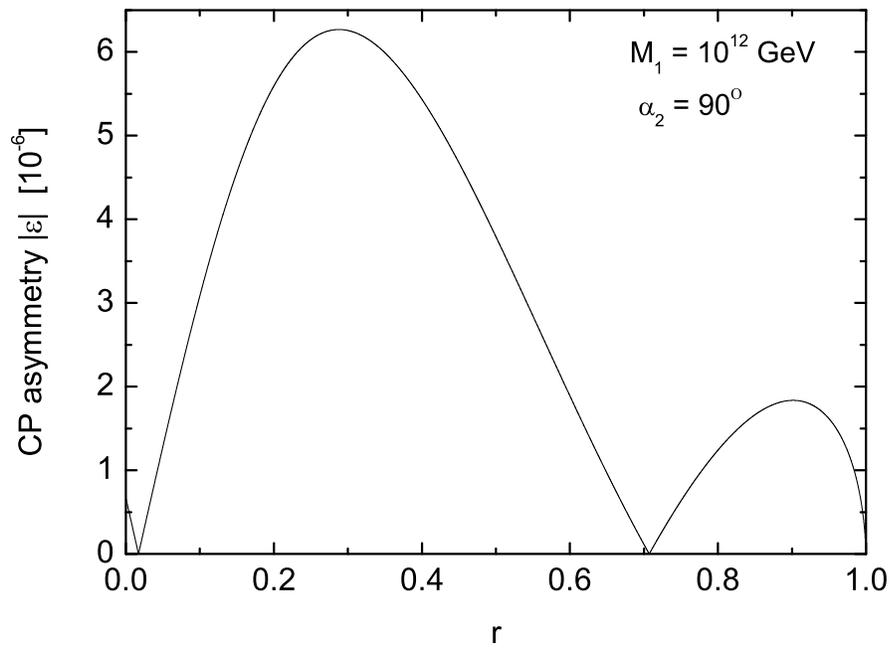}
  \caption{\label{fig:fig3}Absolute value of the CP asymmetry $\epsilon$.}
\end{figure}

Fig.~\ref{fig:fig2} shows the CP asymmetry $\epsilon$ as a function of $r$ with $M_1=10^{12}$GeV and $\alpha_2=90^\circ$. The horizontal line shows $\epsilon=0$. Note that in the small and large $r$ region the CP asymmetry $\epsilon$ becomes negative. Thus, the absolute value of the CP asymmetry has several minimums and maximums as shown in Fig.~\ref{fig:fig3}. According to the relation of $\eta_B \propto Y_L \propto d\vert \epsilon \vert$, if the dilution factor $d$ increases or decreases simply with $r$, we can expect that the baryon-photon ration $\eta_B$ also has several minimums and maximums.

\begin{figure}
  \includegraphics{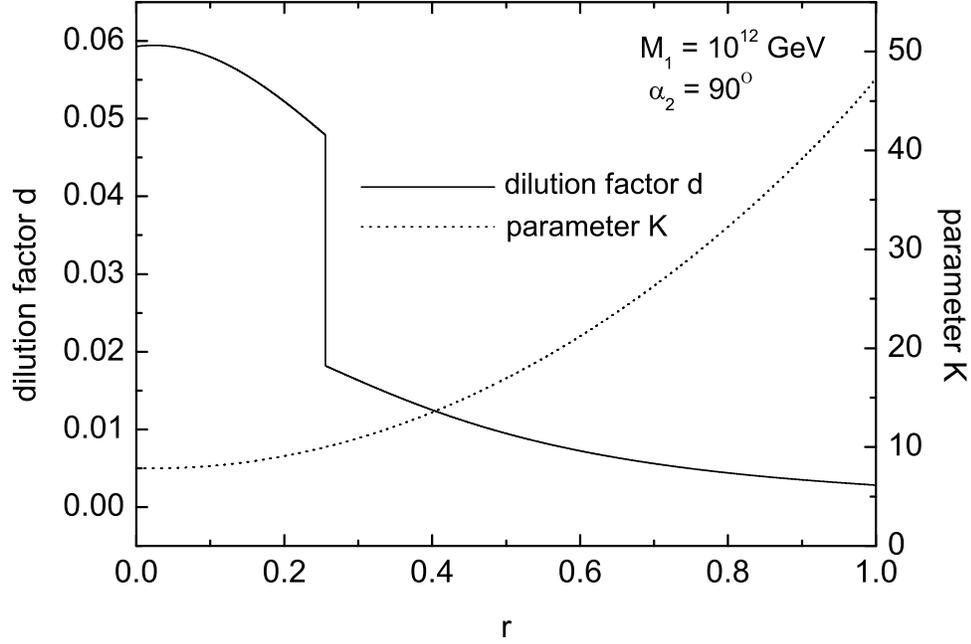}
  \caption{\label{fig:fig4}Dilution factor $d$ as a function of $r$. The parameter $K$ is shown in Eq.~(\ref{Eq:K}).}
\end{figure}

\begin{figure}
  \includegraphics{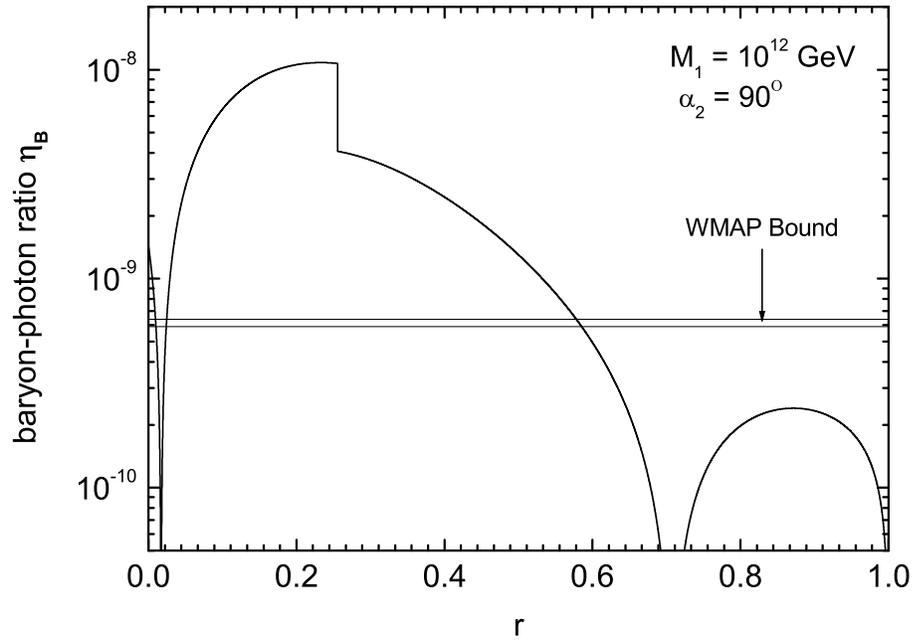}
  \caption{\label{fig:fig5}Baryon-photon ratio $\eta_B$ as a function of $r$.}
\end{figure}

Indeed, the dilution factor $d$ decreases with $r$ as shown in Fig.~\ref{fig:fig4}. The discontinuity at $r\sim 0.26$ arises as long as we use the Nielsen's approximations in Eq.~(\ref{Eq:dilution}) to estimate the dilution factor $d$. Also, it has been pointed out that in some cases the Nielsen's approximations could underestimate the suppression in the baryon asymmetry due to the washout effects \cite{dilutionFactor}. However, the Nielsen's approximations is quite simple, so that we adopt their approximations to simplify our analysis. 

As mentioned above, shown in Fig.~\ref{fig:fig5} is the first main result in this paper. In this figure, we plot the baryon-photon ratio $\eta_B$ as a function $r$ with $M_1=10^{12}$GeV and $\alpha_2=90^\circ$. The reasons of all the behavior of the curve are theoretically understood now. The minimums/maximums and the discontinuity at $r\sim 0.26$ in this figure is caused by the minimums/maximums of the CP asymmetry in Fig.~\ref{fig:fig3} and by the discontinuity of the dilution factor in Fig.~\ref{fig:fig4}, respectively. From the figure, we see the baryon asymmetry significantly varies with $r$, equivalently with $b_2$, e.g., the baryon asymmetry can be enhanced or suppressed at the particular points of $r$. It suggests that we can expect the presence of more rich phenomena, such as an enhanced leptogenesis \cite{enhancedLeptogenesis}, in the $b_2\neq 0$ case. 

\begin{figure}
  \includegraphics{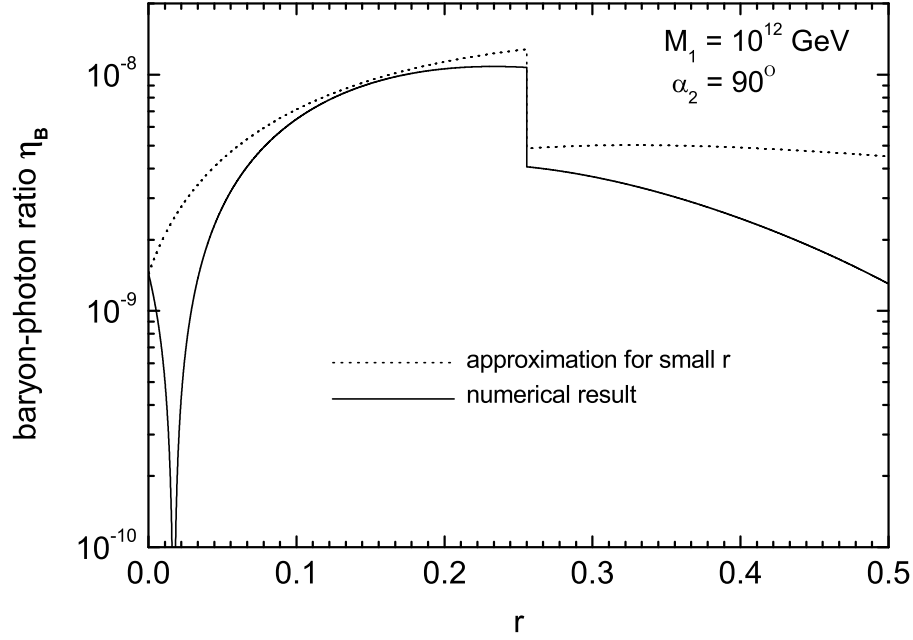}
  \caption{\label{fig:fig6}Approximation for small $r$.}
\end{figure}

\begin{figure}
  \includegraphics{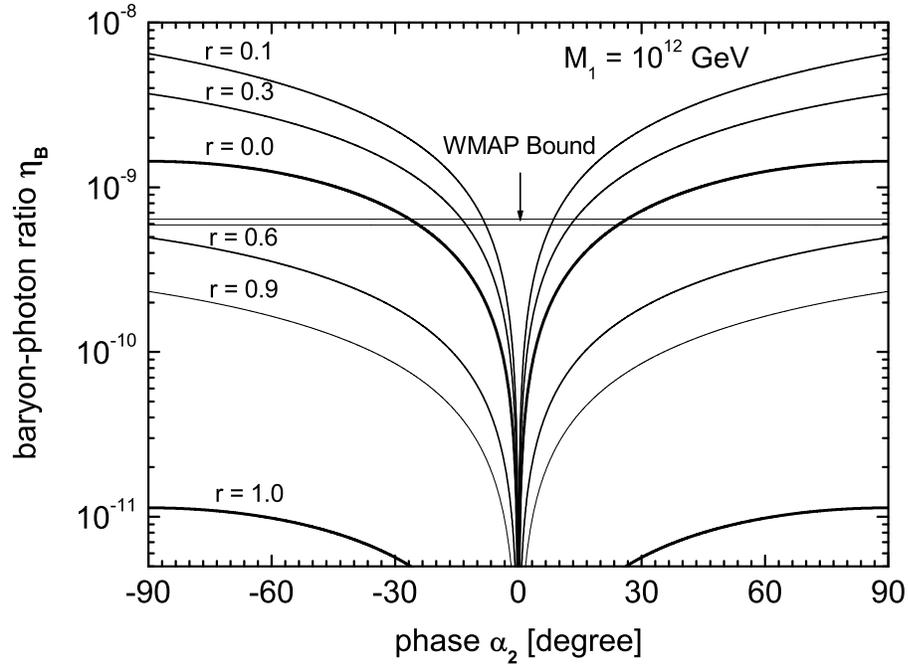}
  \caption{\label{fig:fig7}Baryon-photon ratio $\eta_B$ as a function of the Majorana phase $\alpha_2$ with various values of $r$.}
\end{figure}

To check the consistency of our numerical result, we have obtained the approximate formula of the CP asymmetry parameter in Eq.~(\ref{Eq:epsilon_tild}) up to the first order of $r$ as follows
\begin{eqnarray}
\tilde{\epsilon}=\frac{1}{\vert M_{22}\vert}\frac{\textrm{Im}[A^2] + 2rC\textrm{Im}[A]}{B - 2r\textrm{Re}[A]} + \mathcal{O}(r^2),
\end{eqnarray}
where
\begin{eqnarray}
A&=&\sigma_1 M_{12}^\ast Y +  \sigma_3 M_{23}^\ast Z,
\nonumber\\
B&=&\vert M_{12} \vert^2 + \vert M_{22} \vert^2 + \vert M_{23} \vert^2,
\nonumber \\
C &=& B - \vert Y \vert^2 - \vert Z \vert^2,
\nonumber\\
Y &=& \sqrt{M_{22} M_{11} - M_{12}^2},
\nonumber \\
Z &=& \sqrt{M_{22}M_{33}-M_{23}^2}.
\end{eqnarray}
Fig.~\ref{fig:fig6} shows that the full numerical result is well explained by the approximate formula for small $r$ without the pole at $r \sim 0.16$. It enhances the validity of our numerical result.

Up to now, we have fixed the value of the Majorana phase $\alpha_2$ to $90^\circ$. Finally, we relax this fixing, however we still assume $M_1=10^{12}$GeV. Shown in Fig.~\ref{fig:fig7} is the second main result in this paper. In this figure, we show the baryon-photon ratio $\eta_B$ as a function of the Majorana phase $\alpha_2$ with various value of $r$. Recall that the Majorana phase $\alpha_2$ is the only leptonic CP violation source in our analysis and the baryon-photon ratio $\eta_B$ can be translated into the leptonic CP asymmetry. Thus, it is a natural consequence that the baryon-photon ratio tends to vanish if the Majorana phase $\alpha_2$ goes to zero. Also, Fig.~\ref{fig:fig7} shows that the maximal Majorana CP phase $\alpha_2 = 90^\circ$ is the one of the ingredients which is necessary to generate the maximal baryon asymmetry in the Universe. 

\section{\label{sec:Summary}Summary}
Summarizing our discussion, we have studied the generation of the baryon asymmetry in the Universe in the framework of the minimal seesaw model with tri/bi-maximal mixing. Usually, at least one of the elements in the Dirac mass matrix $m_D$ is fixed to be zero in order to reconstruct all elements in the $m_D$ analytically. For example, $b_2$ element is fixed to zero. In this paper, we have taken non-zero $b_2$. From numerical analysis, we have pointed out that the absolute value of the CP asymmetry $\epsilon$ has several minimums and maximums in the range of $0 \le b_2 \le \sqrt{M_{22}}$. Thus the baryon asymmetry $\eta_B \propto \epsilon$ significantly varies with $b_2$ and one can naturally expect that more rich phenomena, such as an enhanced leptogenesis, are hidden in the $b_2 \neq 0$ case.

\begin{acknowledgments}
The author would like to thank Professor M. Yasu\`{e} for useful comments and careful reading of this paper.
\end{acknowledgments}


\newpage

\end{document}